%% file: Template.tex
\documentclass{article}
\usepackage{spconf,amsmath,graphicx}
\usepackage{multirow}
\title{VADOI: Voice-Activity-Detection Overlapping Inference for End-to-end Long-form Speech Recognition}
\name{Jinhan Wang$^1$, Xiaosu Tong$^2$, Jinxi Guo$^2$, Di He$^2$, Roland Maas$^2$ }
\address{  $^1$Department of Electrical and Computer Engineering, University of California Los Angeles, USA \\
  $^2$Amazon Alexa, USA}
\begin{document}
%\ninept
%
\maketitle
\begin{abstract}
While end-to-end models have shown great success on the Automatic Speech Recognition task, performance degrades severely when target sentences are long-form. The previous proposed methods, (partial) overlapping inference are shown to be effective on long-form decoding. For both methods, word error rate (WER) decreases monotonically when overlapping percentage decreases. Setting aside computational cost, the setup with 50\% overlapping during inference can achieve the best performance. However, a lower overlapping percentage has an advantage of fast inference speed. In this paper, we first conduct comprehensive experiments comparing overlapping inference and partial overlapping inference with various configurations. 
We then propose Voice-Activity-Detection Overlapping Inference to provide a trade-off between WER and computation cost. Results show that the proposed method can achieve a 20\% relative computation cost reduction on Librispeech and Microsoft Speech Language Translation long-form corpus while maintaining the WER performance when comparing to the best performing overlapping inference algorithm. We also propose Soft-Match to compensate for similar words mis-aligned problem.
\end{abstract}
\begin{keywords}
Long-form Speech Recognition, Overlapping Inference, VAD
\end{keywords}
\section{Introduction}
\label{sec:intro}
End-to-end (E2E) models have shown great performance on the Automatic Speech Recognition (ASR) task, and give significant improvement over conventional Hidden Markov Model (HMM) systems \cite{chiu2018state, graves2014towards, dong2018speech}. Compared to conventional systems, E2E systems also yield less memory usage because they maps acoustic input into transcription with a single model. Some of the most popular E2E ASR approaches include Connectionist temporal classification (CTC) \cite{graves2006connectionist, higuchi2020mask}, Recurrent Neural Network Transducer (RNN-T) \cite{graves2012sequence, narayanan2021cascaded}, and attention-based encoder-decoder model \cite{bahdanau2014neural, inaguma2020enhancing}. 

%\cite{yuan2021decoupling}
%jain2019rnn,
%soltau2016neural, 
%li2019towards, 
%  zhang2017very,

However, researchers have shown that E2E models trained on short training segments do not perform well when decoding long-form speech because of domain-mismatch between training and inference phases \cite{chiu2019comparison, chiu2021rnn, narayanan2019recognizing,li2021long}. One solution is to include more long-form audio data during training. But collecting long-form data is time-consuming, and not feasible in model training because of GPU memory restriction. 

Some methods have been introduced to solve the long-form speech recognition problem for E2E models. By incorporating datasets from various domains, models can have better generality regarding various lengths \cite{narayanan2019recognizing}. 
%However, it still falls into memory issue when training on long-form utterances. Therefore, to resolve the hard constraint problem,
Random State Sampling and Passing are proposed to simulate long-form characteristics in the training domain \cite{narayanan2019recognizing}.  %Instead of resetting long-short-term memory (LSTM) state after processing each utterance, initial state are propagated without reset for several steps to let LSTM state convey information from previous states. But, because gradients are not passed back to previous states, this algorithm is not exactly equivalent to simulate training on long-form utterance. 
Monotonic attention is also applied to improve the generality of long-form speech recognition performance by restricting the decoder to focus on a subsequence of encoder states \cite{chiu2021rnn}.
%Therefore, for attention-based models, problem of linear computation cost in sequence length is mitigated compared with soft attention. However, researchers have also proved that manipulating attention behavior is not sufficient to solve long-form problem without the help with proposed overlapping inference (OI).   
Relative Positional Embedding is proposed in order to increase transformer generalization for long-form speech \cite{zhou2019improving}. 

The problem can also be mitigated in the inference stage. In \cite{chiu2018state}, overlapping inference (OI) is proposed. Utterances are decoded after chopping them into short overlapped segments. Segmental hypotheses are aligned and concatenated. However, the non-overlapped region is not tackled properly, and makes OI not applicable with low overlapping percentage. An extension of OI, dynamic overlapping inference is proposed to relax overlapping percentage constraint, by directly aligning frame-level hypothesis from RNN-T \cite{chiu2021rnn}. But there is a large performance gap using a shorter interval. Thus, partial overlapping inference (POI) is introduced to solve the non-overlapped region problem using different margin conditions \cite{kang2021partially}. But POI still degrades recognition accuracy due to the lack of common words under lower overlapping percentages. 

Previous work demonstrated that a higher overlapping percentage yields better performance, but introduce more computation cost. This motivates us to search for a solution that can suppress computation cost by using lower overlapping percentage, like 30\%, but maintain equivalent performance as using 50\%. In this work, we propose Voice-Activity-Detection Overlapping Inference (VADOI), to detect the boundary of the overlapped segment in streaming manner. By obtaining better recognition results at windows boundaries, we can mitigate alignment confusion by introducing more common words. Results show that we can achieve equivalent performance as using 50\% overlapping percentage, with 20\% computation cost reduction on long-form datasets simulated from Microsoft Language Speech Translation (MSLT) \cite{federmann2016microsoft} and Librispeech \cite{panayotov2015librispeech}. We also modify the fixed substitution cost and matching reward into Soft-Match to compensate for mismatch between similar but not identical words. However, because Soft-Match aims to solve edge cases, performance doesn't improve significantly. 

The remainder of this paper is organized as follows: Section 2 introduces OI and POI. Proposed VADOI, Soft-Match are given in Section 3. Experimental results and discussions are reported in Section 4. Section 5 concludes the paper.
%These guidelines include complete descriptions of the fonts, spacing, and
%related information for producing your proceedings manuscripts. Please follow
%them and if you have any questions, direct them to Conference Management
%Services, Inc.: Phone +1-979-846-6800 or email
%to \\\texttt{papers@2021.ieeeicassp.org}.

\section{(Partial) Overlapping Inference}
\label{sec:OIPOI}

In this section, OI and POI are briefly introduced. To solve long-form speech recognition problem in the decoding stage, the naive approach is to concatenate non-overlapped segmental decoded results. However, when naively chopping long-form utterances without constraint, words at boundaries are highly likely to be chopped apart and cause distortion.

\subsection{Overlapping Inference}
\label{sssec:OI}
For OI, decoded results from each pair of overlapped consecutive windows are aligned through dynamic programming. A series of word pairs are concatenated through the tie-breaking algorithm \cite{chiu2019comparison}. The target of alignment is to minimize word error rate (WER) between decoded results of consecutive segments, where previous sentence behaves as groundtruth and next sentence as prediction. There is no surprise that the non-overlapped region will introduce external insertion and deletion errors. This problem doesn't affect alignment performance when the overlapping percentage is high because enough common words can provide sufficient matching reward for good alignment. However, performance degrades dramatically, or even fails when the overlapping percentage is low.

\subsection{Partial Overlapping Inference}
\label{sssec:POI}
To relax the constraint of high overlapping percentage, POI is proposed to suppress external insertion and deletion costs. Compared with OI, marginal deletion cost and insertion cost are set to be 0. To prevent concatenating two overlapped sequences end-to-end, POI sets negative matching reward to encourage overlapping alignment. Aside from manipulating marginal conditions, POI can choose to do alignment in char-level. However, even though computation cost reduces when using lower overlapping percentage, WER degrades monotonically. For algorithm details, please refer to \cite{kang2021partially}.

\section{Proposed Methods}
\label{sec:Propose}
\subsection{VAD Overlapping Inference}
\label{sssec:VADOI}
%In Section 2, we state that performance degrades as overlapping percentage becomes lower for POI, as a tradeoff against faster inference.
As mentioned in Sec. 2, for POI, there is a zero-sum trade-off between computational cost and ASR performance through overlapping inference.
 Error analysis shows that, the cause of WER degradation is lacking shared words across segments. Alignment only works well when there are enough matching rewards from common words between consecutive segments. This condition can be achieved under large overlapping percentage. Although some words are recognized differently in different segments, the number of common words still prevail. While, for low overlapping percentage, fewer words can provide matching reward. If some are recognized differently due to boundary distortion, alignment will be severely affected.

\begin{figure}[t]
  \centering
  \includegraphics[width=8cm,height=3.6cm]{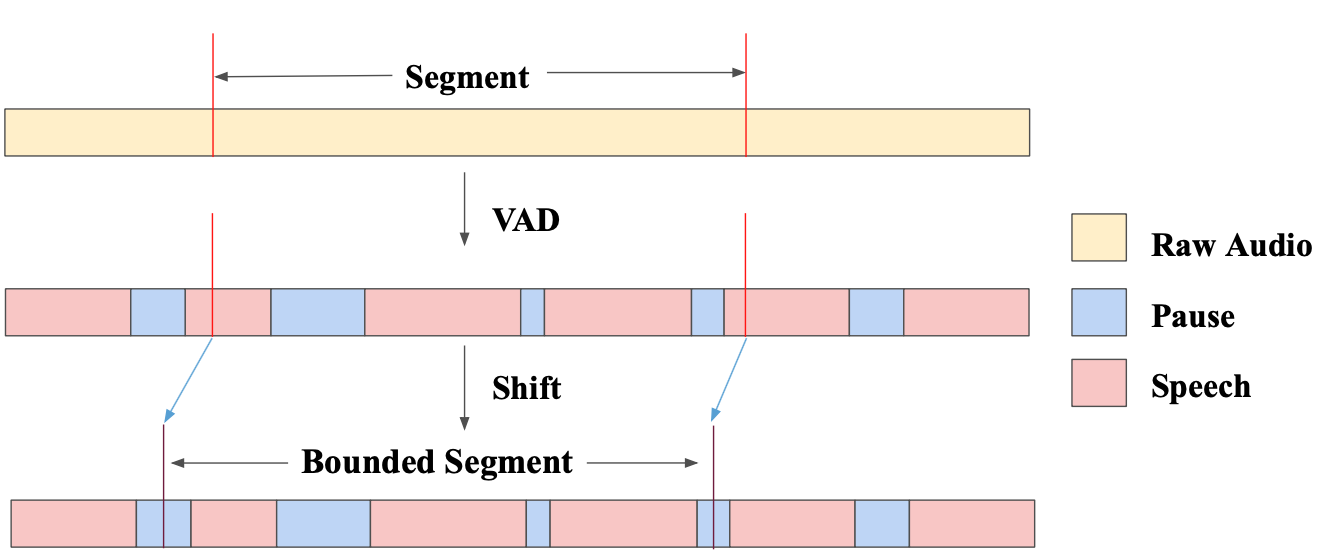}
  \caption{VADOI: Chopping and Shifting}
  \label{fig:VADOI}
\end{figure}

Therefore, VADOI is proposed to prevent chopping in the middle of a word. The diagram of chopping stage of VADOI is shown in Fig.~\ref{fig:VADOI}. After a first-stage segment is generated with fixed segment length and overlapping percentage, a VAD is applied on the segment. The starting frame and the end frame are shifted separately to the middle of the closest long-pause with a desired length. To ensure the existence of an overlapped region, we restrict shifting distance to be shorter than half of the overlapping region length. If a long-pause is not found within this range. The desired length will be cut in half. Searching and shifting will be repeated until a sufficient long-pause is detected. 

For shifting direction, end frame is always shifted to the left to enforce causality. For starting frame, cases are considered separately. When we always shift starting frame to the long-pause on the left, it's possible that some words will appear in three segments where overlapping percentage exceeds 40\%, which will deteriorate alignment. Therefore, starting frame is shifted to the right when overlapping percentage is over 40\%, to prevent triple word-pair, and shifted to the left as shown in Fig.~\ref{fig:VADOI} when overlapping length is below 40\%, to maintain segment length.

Since VAD can be embedded into the model, frame-level VAD results and RNN-T hypothesis can be obtained simultaneously, computation cost for VAD can be ignored.

\subsection{Soft-Match}
\label{sssec:SM}
For OI and POI alignment, the operation cost related to substitution and matching is obtained using the function:
\begin{equation}
\textstyle
e_{sub} = \left\{\begin{matrix}
w_{match} & if \: d_{i}\left ( j \right ) = d_{i+1}\left ( k \right )\\ 
w_{sub} & else
\end{matrix}\right.
\end{equation}
where $d_{i}\left(j\right)$ denotes the $j$-th word in the $i$-th segment, $w_{sub}$ and $w_{match}$ are substitution cost and matching reward, respectively, and $e_{sub}$ is the operation cost \cite{kang2021partially}. 
A substitution error is omitted no matter how similar two words are. Since two words omitted from the same acoustic feature are expected be aligned, Equation (1)'s condition is too strong. To relax the constraint, Soft-Match function is defined as:
\begin{equation}
\textstyle
e_{sub} = CER(d_{i}(j), d_{i+1}(k)) \cdot (w_{sub} - w_{match}) + w_{match}
\end{equation}
The Character Error Rate (CER) between words $d_{i}\left(j\right)$ and $d_{i+1}\left(k\right)$ is projected into a continuous number within range between $w_{sub}$ and $w_{match}$. Then similar word-pair can contribute smaller operation cost to encourage correct alignment.

\section{Experiments}
\label{sec:typestyle}

The RNN-T model is trained on 59k hours of mixed public dataset including Librispeech training data \cite{panayotov2015librispeech}, Switchboard \cite{godfrey1992switchboard}, Fisher \cite{cieri2004fisher}, and so on\footnote{https://github.com/kingformatty/Long-Form-RNN-T-Datasets.git.}, with the Tensorflow framework \cite{abadi2016tensorflow}. 64 dimension Log-filterbank Energy(LFBE) is used and SpecAugment \cite{park2019specaugment} is applied for model training. The encoder consists of 8x1024 LSTM \cite{hochreiter1997long} layers with LayerNorm enabled, and 2x16 single view frequency-LSTM (FLSTM) \cite{li2015lstm} with window size as 8 and stride as 2. The decoder is constructed with 2x1024 LSTM layers. The joint network is a feed-forward structure with single tanh layer. The FastEmit lambda is set to be 0.005 \cite{yu2021fastemit}. All results are obtained without any second pass rescoring or shallow fusion language model.  
%(introduce model we use)
%\input{LaTeX/Tables/Table1}
\subsection{Simulate Long-form Dataset}
\label{sssec:Dataset}
Experiments are conducted on two long-form datasets, MSLT-long and Lib-long, which are simulated from MSLT test set \cite{federmann2016microsoft} and Librispeech test-clean set \cite{panayotov2015librispeech} respectively. For MSLT, we first sort all utterances by length. The long-form test set is simulated by choosing one utterance with duration longer than the median and one shorter than the median repeatedly. Concatenation stops once the length of the concatenated utterance exceeds 120 seconds. For Librispeech, we just simply concatenate utterances from the same speaker. Concatenation stops at 120 seconds. The average length of MSLT-long dataset is 121 seconds and standard deviation as 5 seconds, and 120 seconds and 3.8 seconds for Lib-long. 
%In the following context, MSLT stands for long-form dataset simulated from MSLT dataset, and Lib stands for long-form dataset simulated from Librispeech test-clean. 
%(introduce how we generate long-form dataset without loss of generality)
%We first develop an HMM-DNN hybrid system with BLSTM modelling as our baseline. In addition to sequence-wise model training, we experiment with chunk-wise training using a chunk size of 300 frames, and left and right context chunks of 10 frames each. The table shows that chunk-wise training improves the performance over sequence-wise training mechanism by 5.68\% and 11.55\% in relative WER for the development and evaluation datasets, respectively. The official baseline is also included in the table.
\input{Tables/Table1}

\subsection{(Partial) Overlapping Inference}
\label{sssec:OIPOIexp}
Performance of Baseline, Naive-approach, OI and POI under various configurations on MSLT-long dataset are reported in Table.\ref{tab:ovl}. For the Baseline system, long-form utterances are decoded directly. For the Naive-approach, long-form utterances are chopped into 12 seconds segments without overlapping. Long-form results are obtained by concatenating segments results in order. Naive-approach is marked as 0\% in following tables to represent that there is no overlap. For OI, $w_{del}$, $w_{ins}$, $w_{sub}$, $w_{match}$ are empirically set to 1, 1, 1, 0. For POI, $w_{del}$, $w_{ins}$, $w_{sub}$, $w_{match}$ are set to 2, 2, 1, -2.
%We also report results with different alignment units, word and char, and three different overlapping percentages, 50\%, 30\% and 15\% for OI and POI. % on MSLT and Lib datasets. 

Aside from WER comparison, decoding time and overlapping inference time is also reported. Decoding time represents how many folds is needed to decode compared with Baseline (1T represents the baseline runtime). It also represents the ratio of the number of generated segments compared with Naive-approach (T). Overlapping inference time represents absolute duration for OI-based algorithm to do alignment and concatenation in sec/utt. Results from Table.~\ref{tab:ovl} show that by incorporating OI and POI, performances outperform Baseline and Naive-approach for most cases. Regarding POI and OI, POI always performs better than OI, especially for 30\% and 15\% cases. Since POI sets better margin conditions, non-overlapped regions are handled well. It also shows that word-level alignment almost always outperforms char-level one. Since for char-level alignment, the omitted word might not be in the vocabulary, which introduces extra substitution errors. It's worth noting, OI fails when applying char-level alignment. Since char-level alignment introduces many more non-overlapped instances than word-level one, OI is therefore not compatible with char-level alignment by nature. It can also be observed that, even POI yields good improvement over the Baseline and the Naive-approach when overlapping percentage is low, there is still monotonic performance degradation as overlapping percentage decreases. Regarding decoding time and overlapping inference duration, intuitively, setting larger overlapping percentage will generates more segments which will increase the decoding time by 1.87$\times$ (50\%), 1.37$\times$ (30\%) and 1.16$\times$ (15\%) compared with the Baseline and the Naive-Approach. Char-level alignment takes significant amount of extra time to do alignment and concatenation. Because when we switch alignment unit from word to char, the dynamic graph size scales up exponentially. Searching a extremely large dynamic graph is not feasible in real-world use-case. Based on our findings here, word-level POI is used as the default configuration in the following experiments. We observed similar outcome on Lib-long dataset, and therefore dropped the result in the interest of saving article space.

\subsection{VAD Overlapping Inference}
\label{sssec:VADOIexp}
We employed a statistical-based VAD \cite{sohn1999statistical}. The initial duration of long-pause is empirically set to 0.1 second. Results of VADOI of MSLT-long are reported in Table.\ref{tab:VADOI-MSLT} and Lib-long in Table.\ref{tab:VADOI-Lib}.
Because end frame is always shifted to the left, incorporating VAD will generate slightly more segments under the same overlapping percentages.

For 50\% overlapping percentage, performance of VADOI is slightly worse on both datasets. We believe it is because 50\% overlapping percentage already introduces sufficient common words for good alignment. More segments might not improve alignment performance but rather increase the risk of choosing wrong words. For 30\% overlapping percentage, performance is improved from 13.27\% to 13.02\% on MSLT-long and 6.79\% to 6.58\% on Lib-long, which is equivalent to performance of 50\% on MSLT-long, and only 1.3\% relative WER degradation on Lib-long, but with 19.79\% and 20\% less computation cost respectively. Other computation costs are ignored, including dynamic programming and concatenation, since their computational loads is irrelevant to inferring the RNN-T model. It turns out that for 30\% overlapping percentage, VADOI improves WER by chopping segments more intelligently and cause less confusion for alignment when common words do not prevail. For 15\%, VADOI still outperforms POI. But VADOI on 15\% doesn't achieve equivalent results as using 50\% due to extremely lacking of common words. We also tried 7\% and lower overlapping percentage. But results show that 7\% POI starts to perform worse than applying VAD in Naive-approach.  

\input{Tables/Table2}
\input{Tables/Table3}
\input{Tables/Table4}

\subsection{Soft-Match}
\label{sssec:SMAOIexp}
Table.\ref{tab:SMAOI} reports results when applying Soft-Match on VADOI.
The table shows that by applying Soft-Match in alignment, consistent yet limited improvement is observed for all trials on two datasets. We suspect the reason for lacking strong outcome might be that the problem expected to be solved by Soft-Match does not prevail in our case. However, Soft-Match do not introduce side effect to the performance and can actually handle mis-aligned similar words problem efficiently. 

\section{Summary and Conclusion}
\label{sec:Summary}

In this work, we made a comprehensive comparison of OI and POI with various configurations. Results show that the word-level POI algorithm yields the best performance of all, setting intelligent margin conditions. We also show that by incorporating VAD into POI can efficiently reduce computation cost. With our methods, we have obtained equivalent WER as using 50\% overlapping percentage while only overlapping the segments by 30\% on both MSLT-Long and Lib-long dataset, achieving a 20\% computational cost reduction We also propose a novel Soft-Match mechanism to project the operation cost of substitution and matching into continuous space to compensate for mis-aligned similar words. For future work,we plan to train a better Neural-Network based VAD to replace the naive statistical VAD. We will also investigate how to use acoustic features to estimate optimal overlapping percentage and segmentation length.

% The best one can achieve equivalent performance as using 50\% overlapping percentage, but with around 20\% less computation cost using 30\% on both MSLT-long and Lib-long datasets.
\vfill\pagebreak

%\section{REFERENCES}
%\label{sec:refs}

% References should be produced using the bibtex program from suitable
% BiBTeX files (here: strings, refs, manuals). The IEEEbib.bst bibliography
% style file from IEEE produces unsorted bibliography list.
% -------------------------------------------------------------------------
\bibliographystyle{IEEE}
\bibliography{mybib}

\end{document}

%% file: Tables/Table1.tex
\begin{table*}[t]
\centering
  \caption{WER(\%) and Computation Cost on Various Decoding Schemes on MSLT-long (Decoding time means the computation time needed compared to decoding long-utterance directly. Ovl-Inf Time stands for Overlapping Inference Time, represents absolute alignment and concatenation execution duration in unit sec/utt.)}
\label{tab:ovl}
\begin{tabular}{cccccc}

& & \multicolumn{2}{c}{OI} & \multicolumn{2}{c}{POI} \\  
\multicolumn{2}{c}{WER(\%)/Decoding Time/Ovl-Inf Time} & word                  & char                   & word                   & char                   \\ \hline
\multicolumn{2}{c}{Baseline}     & \multicolumn{4}{c}{20.1/T/NA}                                           \\ \hline
%\multicolumn{2}{c}{Naive Approach}                                                                                            & \multicolumn{4}{c}{16.4(T/NA)}                                           \\ \hline
\multirow{3}{*}{Ovl Percentage} & 0\% &
\multicolumn{4}{c}{16.4/T/NA}
        \\ \cline{2-6}
&  50\%                                                                                                     & 13.6/1.87T/0.88         & 17.0/1.87T/20.5         & 13.1/1.87T/0.88          & 13.2/1.87T/20.5          \\ \cline{2-6}
                     &30\%                                                                                                     & 14.9/1.37T/0.64       & 54.5/1.37T/14.86      & 13.3/1.37T/0.64        & 13.2/1.37T/14.86       \\ \cline{2-6}
                     &15\%                                                                                                     & 25.2/1.16T/0.53       & 71.9/1.16T/12.12       & 13.6/1.16T/0.53        & 14.1/1.16T/12.12       \\ 
\end{tabular}
\end{table*}

%% file: Tables/Table2.tex
% Please add the following required packages to your document preamble:
% \usepackage{multirow}
\begin{table}[]
\centering
  \caption{WER(\%) and Decoding Time of VADOI on MSLT-long. }
\label{tab:VADOI-MSLT}

\begin{tabular}{cccc}
Exp                             & VAD & WER(\%) & Decoding Time \\ \hline
\multirow{2}{*}{0\%} & No  & 16.40   & T            \\
                                & Yes & 14.04   & 1.05T            \\ \hline
\multirow{2}{*}{50\%}           & No  & \textbf{13.05}   & \textbf{1.87T}            \\
                                & Yes & 13.07   & 2.14T            \\ \hline
\multirow{2}{*}{30\%}           & No  & 13.27   & 1.37T            \\
                                & Yes & \textbf{13.02}   & \textbf{1.50T}            \\ \hline
\multirow{2}{*}{15\%}           & No  & 13.59   & 1.16T            \\
                                & Yes & 13.27   & 1.25T           
\end{tabular}
\end{table}

%% file: Tables/Table3.tex
% Please add the following required packages to your document preamble:
% \usepackage{multirow}
\begin{table}[h]
\centering
  \caption{WER(\%) and Decoding Time of VADOI on Lib-long.}
\label{tab:VADOI-Lib}

\begin{tabular}{cccc}
Exp                             & VAD & WER(\%) & Decoding Time \\ \hline
\multirow{2}{*}{0\%} & No  & 9.70   & T            \\
                                & Yes & 7.50   & 1.06T            \\ \hline
\multirow{2}{*}{50\%}           & No  & \textbf{6.49}   & \textbf{1.85T}         \\
                                & Yes & 6.62   & 2.11T            \\ \hline
\multirow{2}{*}{30\%}           & No  & 6.79   & 1.36T            \\
                                & Yes & \textbf{6.58}   & \textbf{1.48T}            \\ \hline
\multirow{2}{*}{15\%}           & No  & 7.28   & 1.13T            \\
                                & Yes & 6.67   & 1.23T           
\end{tabular}
\end{table}

%% file: Tables/Table4.tex
%\begin{table}[H]
%\centering
%  \caption{WER(\%) of VADOI with Soft-Match and AOI}
%  \label{tab:SMAOI}
%\begin{tabular}{ccccc}
%MSLT-long   & WER(\%) &  & Lib-long    & WER(\%) \\ \cline{1-2} \cline{4-5} 
%50\%   & 13.07   &  & 50\%   & 6.62    \\
%\hspace{2.8mm}+ AOI  & 13.06   &  &\hspace{2.8mm} + AOI  & 6.62    \\
%\hspace{5.0mm}+ Soft & 12.99   &  & \hspace{5.0mm} + Soft & 6.59    \\ \cline{1-2} \cline{4-5} 
%30\%   & 13.02   &  & 30\%   & 6.58    \\
%\hspace{2.8mm}+ AOI  & 13.00   &  &\hspace{2.8mm} + AOI  & 6.57    \\
%\hspace{5.0mm}+ Soft & 13.02   &  & \hspace{5.0mm}+ Soft & 6.57    \\ \cline{1-2} \cline{4-5} 
%15\%   & 13.27   &  & 15\%   & 6.67    \\
%\hspace{2.8mm}+ AOI  & 13.28   &  &\hspace{2.8mm} + AOI  & 6.67    \\
%\hspace{5.0mm}+ Soft & 13.25   &  & \hspace{5.0mm}+ Soft & 6.63   
%\end{tabular}
%\end{table}

\begin{table}[h]
\centering
  \caption{WER(\%) of VADOI with Soft-Match}
  \label{tab:SMAOI}
\begin{tabular}{ccccc}
MSLT   & WER(\%) &  & Lib    & WER(\%) \\ \cline{1-2} \cline{4-5} 
50\%   & 13.07   &  & 50\%   & 6.62    \\
+ Soft & 12.99   &  & + Soft & 6.59    \\ \cline{1-2} \cline{4-5} 
30\%   & 13.02   &  & 30\%   & 6.58    \\
+ Soft & 13.00   &  & + Soft & 6.57    \\ \cline{1-2} \cline{4-5} 
15\%   & 13.27   &  & 15\%   & 6.67    \\
+ Soft & 13.25   &  & + Soft & 6.63   
\end{tabular}
\end{table}

%% file: Template.bbl
\begin{thebibliography}{10}

\bibitem{chiu2018state}
C.~C. Chiu, T.~N. Sainath, Y.~Wu, R.~Prabhavalkar, P.~Nguyen, Z.~Chen,
  A.~Kannan, R.~J. Weiss, K.~Rao, E.~Gonina, et~al.,
\newblock ``State-of-the-art speech recognition with sequence-to-sequence
  models,''
\newblock in {\em ICASSP}, 2018, pp. 4774--4778.

\bibitem{graves2014towards}
A.~Graves and N.~Jaitly,
\newblock ``Towards end-to-end speech recognition with recurrent neural
  networks,''
\newblock in {\em ICML}. PMLR, 2014, pp. 1764--1772.

\bibitem{dong2018speech}
L.~Dong, S.~Xu, and B.~Xu,
\newblock ``Speech-transformer: a no-recurrence sequence-to-sequence model for
  speech recognition,''
\newblock in {\em ICASSP}, 2018, pp. 5884--5888.

\bibitem{graves2006connectionist}
A.~Graves, S.~Fern{\'a}ndez, F.~Gomez, and J.~Schmidhuber,
\newblock ``Connectionist temporal classification: labelling unsegmented
  sequence data with recurrent neural networks,''
\newblock in {\em Proceedings of the 23rd ICML}, 2006, pp. 369--376.

\bibitem{higuchi2020mask}
Y.~Higuchi, S.~Watanabe, N.~Chen, T.~Ogawa, and T.~Kobayashi,
\newblock ``Mask ctc: Non-autoregressive end-to-end asr with ctc and mask
  predict,''
\newblock {\em arXiv preprint arXiv:2005.08700}, 2020.

\bibitem{graves2012sequence}
A.~Graves,
\newblock ``Sequence transduction with recurrent neural networks,''
\newblock {\em arXiv preprint arXiv:1211.3711}, 2012.

\bibitem{narayanan2021cascaded}
A.~Narayanan, T.~N. Sainath, R.~Pang, J.~Yu, C.~C. Chiu, R.~Prabhavalkar,
  E.~Variani, and T.~Strohman,
\newblock ``Cascaded encoders for unifying streaming and non-streaming asr,''
\newblock in {\em ICASSP}, 2021, pp. 5629--5633.

\bibitem{bahdanau2014neural}
D.~Bahdanau, K.~Cho, and Y.~Bengio,
\newblock ``Neural machine translation by jointly learning to align and
  translate,''
\newblock {\em arXiv preprint arXiv:1409.0473}, 2014.

\bibitem{inaguma2020enhancing}
H.~Inaguma, M.~Mimura, and T.~Kawahara,
\newblock ``Enhancing monotonic multihead attention for streaming asr,''
\newblock {\em arXiv preprint arXiv:2005.09394}, 2020.

\bibitem{chiu2019comparison}
C.~C. Chiu, W.~Han, Y.~Zhang, R.~Pang, S.~Kishchenko, P.~Nguyen, A.~Narayanan,
  H.~Liao, S.~Zhang, A.~Kannan, et~al.,
\newblock ``A comparison of end-to-end models for long-form speech
  recognition,''
\newblock in {\em ASRU}, 2019, pp. 889--896.

\bibitem{chiu2021rnn}
C.~C. Chiu, A.~Narayanan, W.~Han, R.~Prabhavalkar, Y.~Zhang, N.~Jaitly,
  R.~Pang, T.~N. Sainath, P.~Nguyen, L.~Cao, et~al.,
\newblock ``Rnn-t models fail to generalize to out-of-domain audio: Causes and
  solutions,''
\newblock in {\em IEEE SLT}, 2021, pp. 873--880.

\bibitem{narayanan2019recognizing}
A.~Narayanan, R.~Prabhavalkar, C.~C. Chiu, D.~Rybach, T.~N. Sainath, and
  T.~Strohman,
\newblock ``Recognizing long-form speech using streaming end-to-end models,''
\newblock in {\em ASRU}, 2019, pp. 920--927.

\bibitem{li2021long}
M.~Li, S.~Zhou, and B.~Xu,
\newblock ``Long-running speech recognizer: An end-to-end multi-task learning
  framework for online asr and vad,''
\newblock {\em arXiv preprint arXiv:2103.01661}, 2021.

\bibitem{zhou2019improving}
P.~Zhou, R.~Fan, W.~Chen, and J.~Jia,
\newblock ``Improving generalization of transformer for speech recognition with
  parallel schedule sampling and relative positional embedding,''
\newblock {\em arXiv preprint arXiv:1911.00203}, 2019.

\bibitem{kang2021partially}
T.~G. Kang, H.~G. Kim, M.~J. Lee, J.~Lee, and H.~Lee,
\newblock ``Partially overlapped inference for long-form speech recognition,''
\newblock in {\em ICASSP}, 2021, pp. 5989--5993.

\bibitem{federmann2016microsoft}
C.~Federmann and W.~D. Lewis,
\newblock ``Microsoft speech language translation (mslt) corpus: The iwslt 2016
  release for english, french and german,''
\newblock in {\em IWSLT}, 2016.

\bibitem{panayotov2015librispeech}
V.~Panayotov, G.~Chen, D.~Povey, and S.~Khudanpur,
\newblock ``Librispeech: an asr corpus based on public domain audio books,''
\newblock in {\em ICASSP}, 2015, pp. 5206--5210.

\bibitem{godfrey1992switchboard}
J.~J. Godfrey, E.~C. Holliman, and J.~McDaniel,
\newblock ``Switchboard: Telephone speech corpus for research and
  development,''
\newblock in {\em ICASSP}. IEEE Computer Society, 1992, vol.~1, pp. 517--520.

\bibitem{cieri2004fisher}
C.~Cieri, D.~Miller, and K.~Walker,
\newblock ``The fisher corpus: A resource for the next generations of
  speech-to-text.,''
\newblock in {\em LREC}, 2004, vol.~4, pp. 69--71.

\bibitem{abadi2016tensorflow}
M.~Abadi, P.~Barham, J.~Chen, Z.~Chen, A.~Davis, J.~Dean, M.~Devin,
  S.~Ghemawat, G.~Irving, M.~Isard, et~al.,
\newblock ``Tensorflow: A system for large-scale machine learning,''
\newblock in {\em 12th $\{$USENIX$\}$ symposium on operating systems design and
  implementation ($\{$OSDI$\}$ 16)}, 2016, pp. 265--283.

\bibitem{park2019specaugment}
D.~S. Park, W.~Chan, Y.~Zhang, C.~C. Chiu, B.~Zoph, E.~D. Cubuk, and Q.~V. Le,
\newblock ``Specaugment: A simple data augmentation method for automatic speech
  recognition,''
\newblock 2019.

\bibitem{hochreiter1997long}
S.~Hochreiter and J.~Schmidhuber,
\newblock ``Long short-term memory,''
\newblock {\em Neural computation}, vol. 9, no. 8, pp. 1735--1780, 1997.

\bibitem{li2015lstm}
J.~Li, A.~Mohamed, G.~Zweig, and Y.~Gong,
\newblock ``Lstm time and frequency recurrence for automatic speech
  recognition,''
\newblock in {\em ASRU}, 2015, pp. 187--191.

\bibitem{yu2021fastemit}
J.~Yu, C.~C. Chiu, B.~Li, S.Y. Chang, T.~N. Sainath, Y.~He, A.~Narayanan,
  W.~Han, A.~Gulati, Y.~Wu, et~al.,
\newblock ``Fastemit: Low-latency streaming asr with sequence-level emission
  regularization,''
\newblock in {\em ICASSP}, 2021, pp. 6004--6008.

\bibitem{sohn1999statistical}
J.~Sohn, N.~S. Kim, and W.~Sung,
\newblock ``A statistical model-based voice activity detection,''
\newblock {\em IEEE signal processing letters}, vol. 6, no. 1, pp. 1--3, 1999.

\end{thebibliography}
